\documentstyle[preprint,aps]{revtex}
\begin{document}
\baselineskip 0.7cm
\input{psfig.sty}
\title{    \hfill hep-ph/9612238 (revised March 7. 1997) \\
Breaking  SU3$_f$ Spontaneously Down to Isospin. }
\author{Nils A. T\"ornqvist}
\address{Research Institute for High Energy Physics, SEFT
POB 9, FIN--00014, University of Helsinki, Finland}
\date{7 March 1997}                 
\maketitle
\begin{abstract}
The mechanism where flavour symmetry of is broken spontaneously 
is demonstated for a model
involving nonets of pseudoscalar and vector mesons.
Degenerate bare nonets of
vector  pseudoscalar mesons coupling sufficiently strongly
to each others lead to unstable self-consistency equations, 
such that the SU3$_f$ symmetric spectrum breaks down into a 
stable isospin symmetric mass spectrum similar to the physical
spectrum.
\\
\vskip 0.05cm 
\noindent Pacs numbers:12.39.Ki, 11.30.Hv, 11.30.Qc, 12.15.Ff
\vskip 0.20cm 
\end{abstract}
Conventionally \cite{leut}   
one breaks flavour symmetry by adding effective
non-degenerate quark masses to the QCD Lagrangian,  whereby
the pseudoscalars obtain (small)  masses
and the degeneracy of all flavour  multiplets is split. 
Most of the chiral quark masses are assumed to come from a
short distance regime, where  weak interactions, and the Higgs mechanism
are relevant.

Here I shall  discuss an alternative way of 
breaking  flavour symmetry, which is a  generalisation of  the arguments
given in  a recent paper \cite{NAT3} 
to the physical case of three light flavours.
In that paper I discussed a mechanism
where  a meson mass spectrum, obeying
exact flavour symmetry, the Okubo-Zweig-Iizuka (OZI) rule 
and  general constraints of 
analyticity and unitarity, is unstable 
with respect to quantum loops involving quark pair creation. 
The instability followed from the nonlinear 
self-consistency equations for inverse
 meson propagators containing loops  described diagrammatically 
in Fig. 1.
 
The insight that such a spontaneous flavour symmetry breaking mechanism, rather
than the conventional explicit symmetry breaking through 
quark masses, might be
the true mechanism in the real world, has emerged from earlier studies
of meson masses and widths including meson loops \cite{NAT}, 
where most of the flavour symmetry breaking comes from the 
threshold positions in loops.   

A very important question which arises in this connection is:
Where are the Goldstone degrees of freedom and the Goldstone bosons
expected whenever a symmetry is spontaneously broken? In Ref.~\cite{NAT4}
I argue within  a scalar QCD model  that actually the scalar or longitudinal
confined  gluons are the would-be Goldstone bosons,
not scalar mesons carrying flavour as  was expected in early work\cite{miya}.

In order to make the mechanism for spontaneous flavour symmetry breaking
as transparent as possible I made in  \cite{NAT3} the simplifying
assumption of $C$-degeneracy of multiplets. Thereby one obtains an OZI rule
obeying model, which already for two flavours demonstrates 
why, for the exactly symmetric solution one can have an
unstable situation, - a small symmetry breaking $\Delta m$ in the masses 
of the loop generates an even
bigger symmetry breaking in the output mass spectrum.
A stable self-consistent 
mass spectrum is obtained only  when  the degeneracy of the flavour
multiplet is broken down to an approximately equally spaced spectrum, 
which is ideally mixed and obeys the OZI rule.

The physical mass spectrum, after confinement and chiral 
symmetry breaking in the vacuum, is certainly not $C$-degenerate.
Any confining mechanism should give the quark model rule $C=(-1)^{L+S}$,
and  masses which depend on spin $(S)$ and angular momenta $(L)$. Therefore,
the $C$-degeneracy and consequently  the OZI rule is violated 
in the real world.
In particular, the pseudoscalars ($P$), for which $C=+$,
are not degenerate with
the vectors $(V)$, for which $C=-$, and consequently the OZI rule 
must be violated for loops like  $P\to PV\to P$  
or $V\to PP\to V$.  
  
Here I shall present a model consistent with this fact, i.e. I relax
the $C$-degeneracy condition and apply the arguments to the ground state
nonets of P and V mesons. 
Thereby the model  can approximate the real world for three light flavours,
and if correct it
should generate SU3$_f$ breaking, singlet-octet
splittings, and deviations from 
ideal mixing of the right sign and approximate magnitude. 
The same mechanism
which mixes $s\bar s$ with $u\bar u +d\bar d $ also mixes $u\bar u$ and $d\bar
d$. A priori, it is possible that one finds that
precisely for the linear combinations $u\bar u\pm d\bar d $  of pure isospin
one obtains a stable solution. Then the isospin subgroup is stable i.e., 
{\it not}  broken spontaneously. Here I shall,
in fact,  show that this is what actually happens  in the model presented.

 In the discussion it is useful to 
distinguish four steps of symmetry breaking from the largest symmetry down
to the physical meson spectrum (Fig.2): 
\medskip

(i) The formation of an SU3$_f$ symmetric, but unstable,
spectrum after confinement and
spontaneous chiral symmetry breaking of the vacuum.
Here one allows for explicit vector-pseudoscalar splitting. 
The  pseudoscalar octet $P8$ is composed of  
massless Goldstone bosons, while
 $P1$, $V1$ and $V8$ must at the same time be massive and 
nondegenerate in order to have a spectrum  self-consistent 
with meson loops.

(ii) One generates mass to the pseudoscalar octet, 
but the members remain degenerate. 
The spectrum is thus still SU3$_f$ symmetric 
although chiral symmetry is broken.

(iii) The SU3$_f$ spectrum is broken down to an isospin symmetric
spectum. Conventionally this breaking is 
attributed to the $s-d$ quark mass splitting,
but in this paper I shall instead argue that
the situation after (ii) is unstable, and a spontaneous
symmetry breaking mechanism  gives a similar result in the spectrum.

(iv) Finally a small conventional explicit isospin breaking from
electroweak interactions splits the  isospin multiplets.

\medskip


The three first steps are pictured in Fig. 2. 

The physical flavourless mesons are mixtures of 
$u\bar u$, $d\bar d$ and $s\bar s$ states
determined by an orthogonal $3\times 3$  mixing matrix $\Omega$.
If the unmixed states $q_i\bar q_i$ 
are described by  flavour matrices $\Lambda^{ii}$ the
physical states are described by  
 $\Lambda^\alpha =\sum_i \Omega_{\alpha i}\Lambda^{ii}$. Here $\Omega$  
mixes the the $u\bar u$ and $d\bar d$ states to $u\bar u\pm d\bar d$
when isospin is exact, and furthermore mixes 
the $u\bar u +d\bar d$ and
$s\bar s$ states by a mixing angle $\delta$, measuring the deviation from
"ideal" mixing ($\delta =\theta^{SU3}+{\rm \arctan}\sqrt 2$). 
This mixing matrix  $\Omega $ will be 
determined by  the self-consistency  equation, which is  obtained 
below for the $3\times 3$ mass matrix
in the inverse propagators, and which is diagonalized by $\Omega$.
 Once $\Omega $ is determined the 
flavour related couplings for a vertex $A^\alpha \to B^\beta C^\gamma$
are given by the general formula
\begin{equation}
g^{\alpha\beta\gamma}_{ABC,S} = 
g_{ABC}{\rm Tr}[\Lambda^\alpha_A\Lambda^\beta_B\Lambda^\gamma_C]_S,
  \label{trace}
\end{equation}   
where $S$ is determined by the charge conjugation quantum numbers
of the three multiplets, $S=C_A C_B C_C$. 
Thus the symmetric ($S=+$, for "D-coupling") or antisymmetric ($S=-$, for
"F-coupling") trace, 
must be taken. Therefore singlet states decouple when
$S=-$, but  couple with twice the strength  compared to $8\to 8+8$
transitions if $S=+$.
Eq.(1) allows for OZI rule violation only through particle mixings in the
meson propagator. With isospin exact Eq.(\ref{trace}) implies that all 
$g$'s for one group of thresholds like $P\to PV$  are determined
by only one mass dependent mixing angle ($\delta$) and one overall 
coupling\footnote{
There are other possible flavour symmetric, but OZI rule violating  couplings,
involving at least one  singlet,
 such as Tr$\Lambda_A$Tr$\Lambda_B$Tr$\Lambda_C$ or
Tr$[\Lambda_A\Lambda_B]$Tr$\Lambda_C$ + permutations. 
For simplicity I  do not here include the  extra parameters involved with
these. They would  shift details of the spectrum through shifting the 
octets and singlets, but the spontaneous generation of "$s$-quark mass"
 would still occur.}.

We can now  construct the inverse propagators  for the pseudoscalars   
including the loop contributions from the $PV$ and $VV$ thresholds: 

\begin{eqnarray}
P^{-1}_{\alpha\alpha '}(s)= m_0^2-s +\frac 1{4\pi}\sum^{9,9}_{\beta ,\gamma=1}
\bigl [ [g_{PPV,-}^{\alpha\beta\gamma}g_{PPV,-}^{\alpha '\beta\gamma}]
&{\rm F}&(s,m^2_{P_\beta},m^2_{V_\gamma},\Lambda)+\label{prop} \\
+\frac 1{4\pi} [g_{P´VV,+}^{\alpha\beta\gamma}g_{PVV,+}^{\alpha '\beta\gamma}]
&{\rm F}&(s,m^2_{V_\beta},m^2_{V_\gamma},\Lambda)\bigr ] \ , \nonumber
\end{eqnarray}
where the  function F (See \cite{NAT3}) 
must contain the unitarity cut and the  phase space 
of one loop diagrams. SU6 predicts that the overall couplings are equal:
$g_{PPV} =g_{PVV}$, a constraint which  shall be used here.

Summing over isospin states, and in order not to have too cumbersome notation
choosing the special case of ideally mixed
($\delta_V=0$) vector mesons and   pseudoscalars  threshold masses 
mixed as in the pure SU3$_f$ frame ($\theta^{SU3}_P=0$) the different
thresholds contribute to the 
$\pi ,\ K$ mass matrix with the following weights composed
of Clebsch-Gordan-like coefficients in the partial sums  $\sum
[g_{ABC}^{\alpha\beta\gamma}g_{ABC}^{\alpha '\beta\gamma }]/(g_{ABC})^2$:
\begin{eqnarray}
\pi &:&\  4\rho\pi + 2K\bar K^* +4\rho\omega_u +2K^*\bar K^* \ ,\label{pion} \\
K &:&\  \frac 3 2 K^*\pi+{\frac 3 2}
 K\rho+\frac 1 2 K\omega_u+K\phi_s+\frac 3 2\eta_8K^* + 
\label{kaon} 
\hskip .2cm  3K^*\rho+ K^*\omega_u +2K^*\phi_s \ . 
\end{eqnarray}
Similarily for the $\eta-\eta '$  mass matrix one has:
\begin{eqnarray}
\lefteqn{
\left(\begin{array}{cc}\eta_{88}&\eta_{81}\\ \eta_{18}&\eta_{11}\end{array}
\right) :  \left(\begin{array}{cc}6&0 \\ 0 & 0 \end{array}\right) 
K\bar K^* +  \nonumber  
\frac 2 3 \left(\begin{array}{cc}1&-\surd 2 \\ -\surd 2 &2\end{array}\right)
(3\rho\rho+\omega_u\omega_u) + 
} \\  
&&{\hskip 2.5cm} 
 \frac 4 3 \left(\begin{array}{cc}0.5&\surd 2 \\ \surd 2
 &4\end{array}\right) K^*\bar K^* +  
 \frac 4 3 \left(\begin{array}{cc}2&\surd 2 \\ \surd 2
 &1\end{array}\right)
\phi_s\phi_s  \ .\label{etas}  
\end{eqnarray}
The notation in (\ref{pion}-\ref{etas}) should here be obvious 
[e.g. $4\pi\rho$ stand for 4F$(s,m^2_\pi,m^2_\rho,\Lambda )] $, 
except for the fact that
$K\bar K^*$ stands for $K\bar K^*+\bar K K^*$ and, more precicely
e.g., $K^*\omega_u$ should be replaced by by 
$(\cos \delta_V+\sqrt 2\sin \delta_V )^2K\omega$, and
2$K^*\phi_s $ should be replaced by by 
$(\sin  \delta_V-\sqrt 2\cos \delta_V )^2K\omega$
etc. This is done in the numerical work through $\Omega$ and the formula 
(\ref{trace}), where in addition $\Omega$ and 
the mixing angles depend on mass, e.g.
the pseudoscalar mixing angle is obtained when diagonalizing Eq.
(\ref{etas}) at the $\eta$ and $\eta '$ mass, and similarily
$\delta_V(m_\omega )$ is not exactly equal to $\delta_V(m_\phi )$.

  For the vector mesons one has similar equations as Eq. (\ref{prop}), involving
$PP$, $PV$, and $VV$ thresholds. There SU6  predicts that these should  be 
added with the relative weights $\frac 1 6,\ \frac 4 6, \ \frac 7 6$, but  
such relations are known to agree rather poorly with experiment. For our
demonstrative 
purpose it is sufficient to add the $V\to PP$ thresholds (for which $S=-$)
and the $V\to PV$ thresholds (for which $S=+$) with equal weights  
such that in fact we have 
$g_{PPV} =g_{PVV}=g_{VPP}=g_{VPV}= g$. Thus for
the vectors we have very similar equations as those of Eq. (\ref{prop}) 
for the pseudoscalars, but with $PP$ replacing $PV$ and $PV$ replacing $VV$.

Now consider the SU3$_f$ limit after step (ii) assuming as above the four
overall couplings to be equal. One can  sum over the
degenerate octets, and $\theta^{SU3}$ must vanish. 
For the octet and singlet pseudoscalar propagators one finds:     

\begin{eqnarray} 
P^{-1}_{P8} (s) = \frac 2 3 \frac {g^2} {4\pi} 
[9&{\rm F}&(s,m^2_{P8},m^2_{V8},\Lambda)+  \nonumber \\
5 &{\rm F}&(s,m^2_{V8},m^2_{V8},\Lambda)+  \label{P8}\\
 4 &{\rm F}&(s,m^2_{V1},m^2_{V8},\Lambda)] +m^2_{0,P}- s\ , \nonumber \\
   P^{-1}_{P1} (s) = \frac 4 3 \frac {g^2}{4\pi}
[8 &{\rm F}&(s,m^2_{V8},m^2_{V8},\Lambda)+ \label{P1} \\
    &{\rm F}&(s,m^2_{V1},m^2_{V1},\Lambda)]+m^2_{0,P}- s \ , \nonumber 
\end{eqnarray}
and similar equations for the vectors as discussed above.

When the bare masses are equal ($m_{0,P}=m_{0,V}=m_0$)
 there exists always a self-consistent SU6
solution with all physical masses equal:
$m^2_{P_\alpha}=m^2_{V_\alpha}= m^2_0+\Delta m^2$ 
where $\Delta m^2 =\frac {g^2}{4\pi}12$F.
All inverse propagators are then the same: 
$P^{-1}_{P8}=P^{-1}_{P1}=P^{-1}_{V8}=
P^{-1}_{V1}$ = $\frac {g^2}{4\pi}12{\rm F}
(s,m_0^2+\Delta  m^2,m_0^2+\Delta m^2,\Lambda)+m_0^2- s$.
In fact, in this limit one has a very similar situation as the one discussed
in  \cite{NAT3}.
There I showed that this solution 
is unstable, and that flavour symmetry will be
broken spontaneously by the loops if $g$ is sufficiently large.
The resulting spectrum of the stable solution, which must be found 
numerically even for the simplest possible model for F,
 obeys approximately the equal spacing rule.
 
But once the vectors are  
heavier than the near massless  pseudoscalars,
 $m_0^V> m_0^P$, the situation is more
complicated. Already when  SU3$_f $  remains exact 
the singlet masses  must be 
different from the octet masses as can be seen from the self consistency 
equations (\ref{P8}-\ref{P1}) and as was anticipated in Fig.1. The singlets
will be heavier since the nearest $PV$ thresholds shift only the octet down.
Then for sufficiently large $g$ 
this unstable SU3$_f $  symmetric configuration will be 
spontaneously broken down in a way which must compromize between the ideally
mixed solution  and the OZI rule violating SU3$_f$ symmetric solution.

Also this solution can of course only be found numerically, 
as is done in Fig.3 for different
values of the cutoff $\Lambda$, and for  $g$ very large.
The unstable SU3$_f$ symmetric solution for the singlet and octet states are
shown as the dashed curves, while the stable flavour symmetry violating masses
are the full drawn curves. To the left are the physical masses indicated.
The  $\rho$ and $\pi$ masses are used as input to fix the two subtraction 
constants $m_{0,P}$ and $m_{0,V}$, while $g$ 
is for simplicity assumed very big.
As can be seen already this crude scheme gives a reasonable spectrum.
The mixing angles $\delta$  are of right sign, although a
little too big compared to experiment for the vectors (when $\Lambda=2$ GeV
$\delta (m_\omega )=18.3^\circ $,  $\delta (m_\phi )=22.6^\circ $)
 and too small for the pseudoscalars 
($\delta (m_\eta )=25.1^\circ $, $\delta (m_{\eta '} )=25.7^\circ $).  


Why does the solution stabilize itself to an isospin symmetric one?
Isospin is {\it not} broken spontaneously, because the linear combinations
$u\bar u\pm d\bar d$ distribute the flavour probabilities $u\bar u$
and $d\bar d$ equally in the physical mesons. Therefore the instability
in terms of the $r$ parameter introduced 
in \cite{NAT3} is decreased by a factor of 3,
since only the loop involving an $s\bar s$ state 
contribute to the numerator of $r$, i.e.:

\begin{equation}
r=\frac{ N_f {\rm F}_{m^2}}{-N_f {\rm F}_s+4\pi/g^2 }  \rightarrow 
\frac{ {\rm F}_ {m^2}}{-3{\rm F}_s +4\pi/g^2} \approx \frac 1 3 < 1 \ 
. \label{istab2}
\end{equation}
where 
${\rm F}_s=\frac {\partial{\rm F}(s,m^2_1,m^2_2,1)}
{\partial s}\Bigr|_{s=m_1^2=m^2_2}$ and 
${\rm F}_{m^2}=\frac {\partial{\rm F}(s,m^2_1,m^2_2,1)}
{\partial m_1^2}\Bigr|_{s=m_1^2=m^2_2}$. 
The value of $r$  decreases from slightly above one  (see Fig. 3 of 
\cite{NAT3})  
to about one third which
is well below the instability limit  $r>1$.
In detail this can be understood by looking at the $K^0-K^+$ mass splitting
and the $u\bar u$ and $d\bar d$ 
loop contributions to the $K^0$ 
and $K^+ $   masses. For $K^+\to (\bar s u)(\bar u
u) \to K^+$ or $K^0\to (\bar s d)(\bar d d)\to K^0$ the $u\bar u$
or $d\bar d$ pairs are replaced by $(u\bar u\pm d\bar d)/\sqrt 2$ after the 
isospin rotation in $\Omega$. The latter distribute the flavour
probabilities democratically and neutralizes the instability created if
$d $   is made heavier than $u$. Therefore the isospin symmetric solution 
is a stable solution. Of course isospin will still be broken by 
electroweak effects. 
Thus to understand the small physical isospin breaking
one needs the conventional explicit symmetry breaking mechanism.
Renormalization effects  are of course also then present, 
but should only slightly  enhance the driving term from electroweak physics.

The whole of SU3$_f$ cannot remain unbroken this way, 
since the P-V splitting can
stabilize only one degree of freedom, not two. With three flavours
the neutral $u\bar u$, $d\bar d$ and $s\bar s$ probabilities cannot
be distributed democratically for all three orthogonal flavourless states.
In particular 
the octet states $u\bar u -d\bar d$ and $u\bar u+ d\bar d -2s\bar s$
contain very different amount of $s\bar s$.

The numerical work in this paper (Fig.3) 
has been done only for demonstration, with no
ambitions to fit the data exactly. 
As input parameters there was only the $\pi $
mass, the $\rho $ mass and the cut off 
$\Lambda$, but still a spectrum not too far
from the physical one was obtained. The essential thing which was  demonstated
was that an effective "strange quark mass" 
can be generated  spontaneously within strong
interaction loops, and that its magnitude is related to the $\rho$ and 
$\pi$ masses.
Numerous improvement of the actual fit to the spectrum could be done by 
including   more thresholds
(such as  those involving P-wave $q\bar q$ mesons), 
detailed fits for $g_{PPV}$, $g_{PVV}$ and singlet  coupling parameters  like 
Tr$\Lambda^A$Tr$[\Lambda^B \Lambda^C]$ mentioned in the footnote with pure 
gluonic contributions, more detailed function for F etc.
                                 
I thank especially Geoffrey
 F. Chew and  Claus Montonen  for  useful discussions.

\eject
\begin{figure}
\psfig{figure=fig1nilst.epsf,width=17cm,height=2.2cm}
\vskip 0.2cm
\caption{ The  self-consistency equations diagrammatically. Iterating 
the equation  gives a sum of multiloop diagrams.}
\end{figure}  
\begin{figure}
\psfig{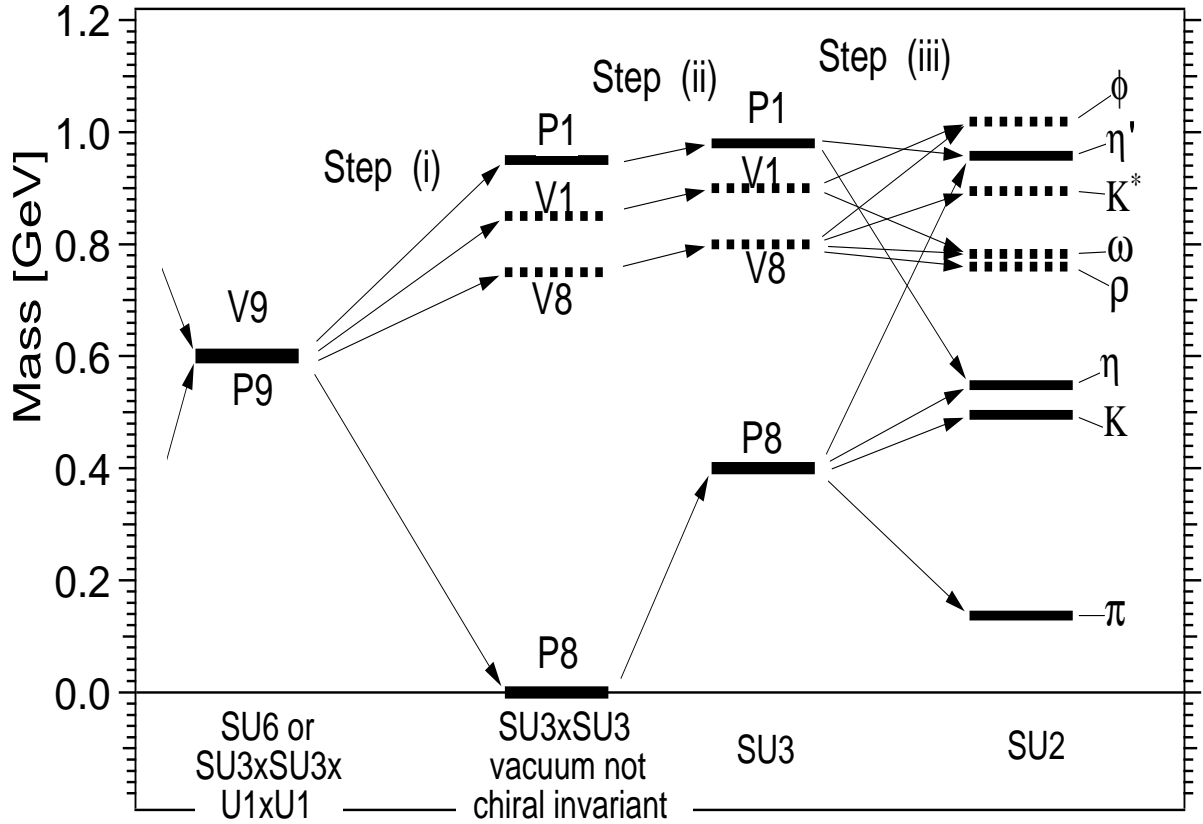}
\caption{ The sequence of steps of 
symmetry breaking from a fully degenerate
SU6 symmetric mass (the value of which depends 
on how the vacuum is defined)
down to the near isospin symmetric physical spectrum (see text). }
\end{figure}                      
\begin{figure}
\psfig{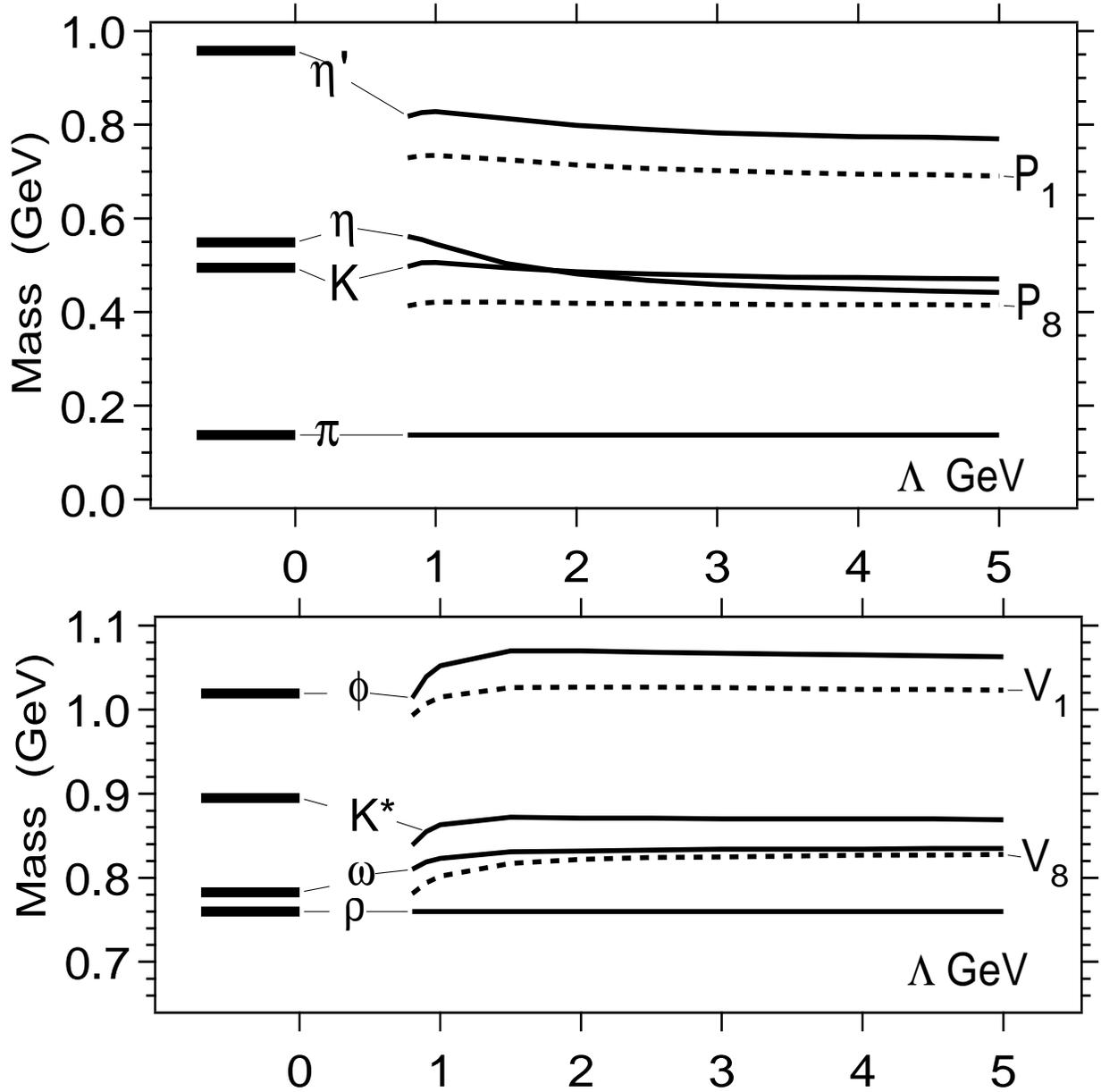}
\caption{ The predicted pseudoscalars and vector meson masses as
functions of $\Lambda$. The $\pi$ and $\rho $   masses are input and fix the
subtraction constants. The dashed lines shows the unstable SU3$_f$ octet and
singlet masses
before the spontaneous breaking (but with the same 
 subtraction constants as for
the stable solution). To the left the experimental masses are shown.
Note in particular the very weak dependence on the cutoff $\Lambda$.   }
\end{figure}

\end{document}